\documentclass[sigconf, screen]{acmart}

\usepackage{graphicx} %
\usepackage{natbib}  %
\usepackage{caption} %
\usepackage{algorithm}
\usepackage{listings}
\usepackage{algorithmic}
\usepackage{amsmath}
\usepackage{booktabs}
\usepackage{multirow}
\usepackage[outdir=./]{epstopdf}
\usepackage{enumitem}
\usepackage{caption}
\usepackage[most]{tcolorbox}
\usepackage{subcaption}
\usepackage{newfloat}
\usepackage{float}
\usepackage{balance}
\usepackage{svg} %
\usepackage[normalem]{ulem}
\usepackage{threeparttable}
\usepackage{multirow} %
\usepackage{xcolor}

\definecolor{darkred}{rgb}{0.5,0,0}
\definecolor{darkblue}{rgb}{0,0,0.5}
\definecolor{darkgreen}{rgb}{84,130,53}

\AtBeginDocument{%
  }

\begin{document}

\title{When Text-as-Vision Meets Semantic IDs in Generative Recommendation: An Empirical Study}

\author{Shutong Qiao}
\affiliation{
  \institution{The University of Queensland}
  \city{Brisbane}
  \country{Australia}
}
\email{shutong.qiao@uq.edu.au}

\author{Wei Yuan}
\affiliation{
  \institution{The University of Queensland}
  \city{Brisbane}
  \country{Australia}
}
\email{w.yuan@uq.edu.au}

\author{Tong Chen}
\affiliation{
  \institution{The University of Queensland}
  \city{Brisbane}
  \country{Australia}
}
\email{tong.chen@uq.edu.au}

\author{Xiangyu Zhao}
\affiliation{
    \institution{City University of Hong Kong}
    \city{Hongkong}
    \country{China}
}
\email{xy.zhao@cityu.edu.hk}

\author{Quoc Viet Hung Nguyen}
\affiliation{
    \institution{Griffith University}
    \city{Gold Coast}
    \country{Australia}
}
\email{henry.nguyen@griffith.edu.au}

\author{Hongzhi Yin}
\affiliation{
    \institution{The University of Queensland}
    \city{Brisbane}
    \country{Australia}
}
\email{h.yin1@uq.edu.au}

\renewcommand{\shortauthors}{Shutong Qiao et al.}
\begin{abstract}
Semantic ID learning is a key interface in Generative Recommendation (GR) models, mapping items to discrete identifiers grounded in side information, most commonly via a pretrained text encoder. However, these text encoders are primarily optimized for well-formed natural language. In real-world recommendation data, item descriptions are often symbolic and attribute-centric, containing numerals, units, and abbreviations. These text encoders can break these signals into fragmented tokens, weakening semantic coherence and distorting relationships among attributes. Worse still, when moving to multimodal GR, relying on standard text encoders introduces an additional obstacle: text and image embeddings often exhibit mismatched geometric structures, making cross-modal fusion less effective and less stable.

In this paper, we revisit representation design for Semantic ID learning by treating text as a visual signal. We conduct a systematic empirical study of OCR-based text representations, obtained by rendering item descriptions into images and encoding them with vision-based OCR models. Experiments across four datasets and two generative backbones show that OCR-text consistently matches or surpasses standard text embeddings for Semantic ID learning in both unimodal and multimodal settings. Furthermore, we find that OCR-based Semantic IDs remain robust under extreme spatial-resolution compression, indicating strong robustness and efficiency in practical deployments. 
\end{abstract}



\maketitle

\section{Introduction}\label{sec_intro}
Recommender systems\cite{burke2011recommender,fu2025forge} play a vital role in helping users discover content of interest in modern information systems. Recently, they have increasingly adopted generative formulations, casting recommendation as autoregressive sequence generation rather than independently scoring candidate items \cite{wang2023generative, hou2025generative, zhai2024actions}, known as Generative Recommendation (GR), which can naturally benefit from advances in sequence modeling. 
A central ingredient enabling this formulation is Semantic IDs: discrete identifiers derived from item side information, rather than conventional, arbitrarily assigned item IDs \cite{rajput2023recommender, singh2024better, yang2024unifying, deng2025onerec}. Compared with conventional item IDs, Semantic IDs inject item semantics as an inductive prior, reducing reliance on memorizing interaction co-occurrence and improving content-based generalization \cite{hou2025generating, zheng2025enhancing}.

Existing Semantic ID learning methods typically construct IDs by first encoding item descriptions into a continuous representation using a pretrained text encoder and then discretizing it into a structured identifier space \cite{ju2025generative, rajput2023recommender}. This text-anchored design implicitly assumes that item descriptions resemble well-formed natural language and can be faithfully modeled by standard text encoders. However, in real-world recommender systems, item descriptions frequently break this assumption: they are often structured and attribute-centric, composed of symbols, numbers, units, and abbreviations rather than fluent sentences. For instance, Amazon product descriptions commonly appear as compact lists such as “Wireless Mouse, 2.4G, 1600DPI, USB Receiver, Black,” where semantics are conveyed primarily through symbolic cues. In such cases, standard text encoders will split these signals into fragmented and even out-of-vocabulary tokens that weaken semantic coherence and distort relationships among attributes, deteriorating the quality of Semantic IDs.
These issues are further exacerbated in multimodal settings, where a common case is that items' side information contains both texts and images. To incorporate both textual and visual information, existing pipelines typically rely on a text encoder and an image encoder from different model families. Because the two encoders are trained with different objectives and inductive biases, and because the modalities capture complementary aspects of item characteristics, the resulting text and image embeddings often reside in mismatched semantic spaces. As a result, obtaining a fused representation for Semantic ID learning necessitates non-trivial and sometimes highly complex cross-modal alignment and fusion mechanisms.

\begin{figure}
    \centering
    \includegraphics[width=0.85\linewidth]{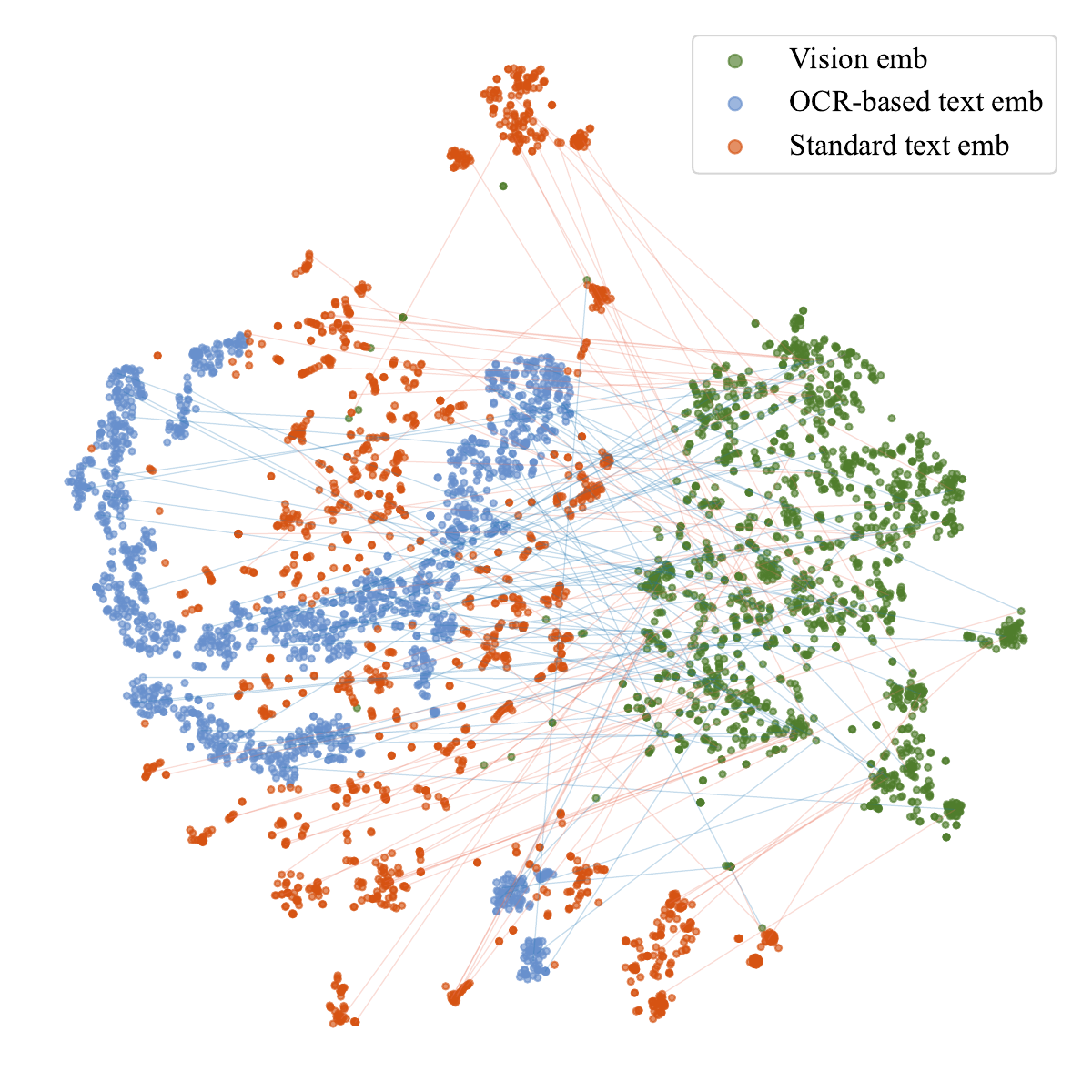}
    \caption{Embedding geometry across modalities. We project three item representations into a shared 2D space: \textbf{Item image emb}, extracted from each item’s photos; OCR-based text emb, extracted by rendering the item’s textual description into an image and encoding it with an OCR model; and Standard text emb, produced by a standard text encoder operating on item’s textual description. Each point denotes an item, and cross-modal lines connect representations of the same item. OCR-text embeddings form smooth manifolds that align closely with image embeddings, whereas text embeddings exhibit higher dispersion and stronger anisotropy.
    }
    \label{fig:emb}
    
\end{figure}

Recent advances in vision–language modeling have introduced text-as-vision \cite{li2025text, xing2025vision, tai2024pixar, xing2025see}, which renders text into images for visual encoding, with OCR-oriented encoders emerging as a more effective and widely adopted solution than generic vision encoders by explicitly capturing linguistic content together with typography and layout~\cite{chai2024dual, cheng2025glyph, li2025text, wei2025deepseek}.
Such OCR-based representations improve token efficiency and robustness, and have become the dominant instantiation of text-as-vision in practical semantic modeling.
From a representation learning perspective, adopting visual text representations reshapes the geometry of the embedding space, as illustrated in Figure~\ref{fig:emb}, which simulates a multimodal feature encoding setting with item image embeddings and textual features extracted using either standard text embeddings or OCR-based visual text embeddings.
In this setting, OCR-based text embeddings form smoother and less anisotropic manifolds than standard text embeddings, while also exhibiting closer geometric alignment with item image embeddings, thereby narrowing the modality gap and reducing the reliance on explicit cross-modal alignment during fusion.
As a result, OCR-based visual text representations enable more stable semantic abstraction in tasks involving discrete tokenization and multimodal integration.

Inspired by these observations, a natural question arises: can symbolic, attribute-centric descriptions be rendered into images and encoded using vision-centric OCR representations to replace standard text embedding methods? Intuitively, this approach may mitigate the issues of fragmented or out-of-vocabulary tokens inherent in traditional text encoders. Moreover, in multimodal settings, OCR-based text representations are expected to align more naturally with item image embeddings, thereby reducing cross-modal misalignment and yielding a more stable and coherent semantic space for Semantic ID construction. To investigate this hypothesis, we conduct a systematic empirical study evaluating OCR-based text representations as an alternative to standard text embeddings for Semantic ID learning in GR.
Specifically, we investigate the following research questions (RQs):
\begin{itemize}
    \item (\textbf{RQ1}) Can OCR-based text representations substitute standard text representations in a unimodal Semantic ID learning setting?
    \item (\textbf{RQ2}) Can OCR-based text representations substitute standard text representations in a multimodal Semantic ID learning setting?
    \item (\textbf{RQ3}) How robust are OCR-based Semantic IDs to variations in OCR encoders and rendering quality?
\end{itemize}

In summary, rather than proposing a new recommendation architecture, our study focuses on providing principled insights into the effect of representation choices on Semantic ID quality and downstream generation performance, and to characterize when and why OCR-based representations help.

\begin{figure}
    \centering
    \includegraphics[width=0.95\linewidth]{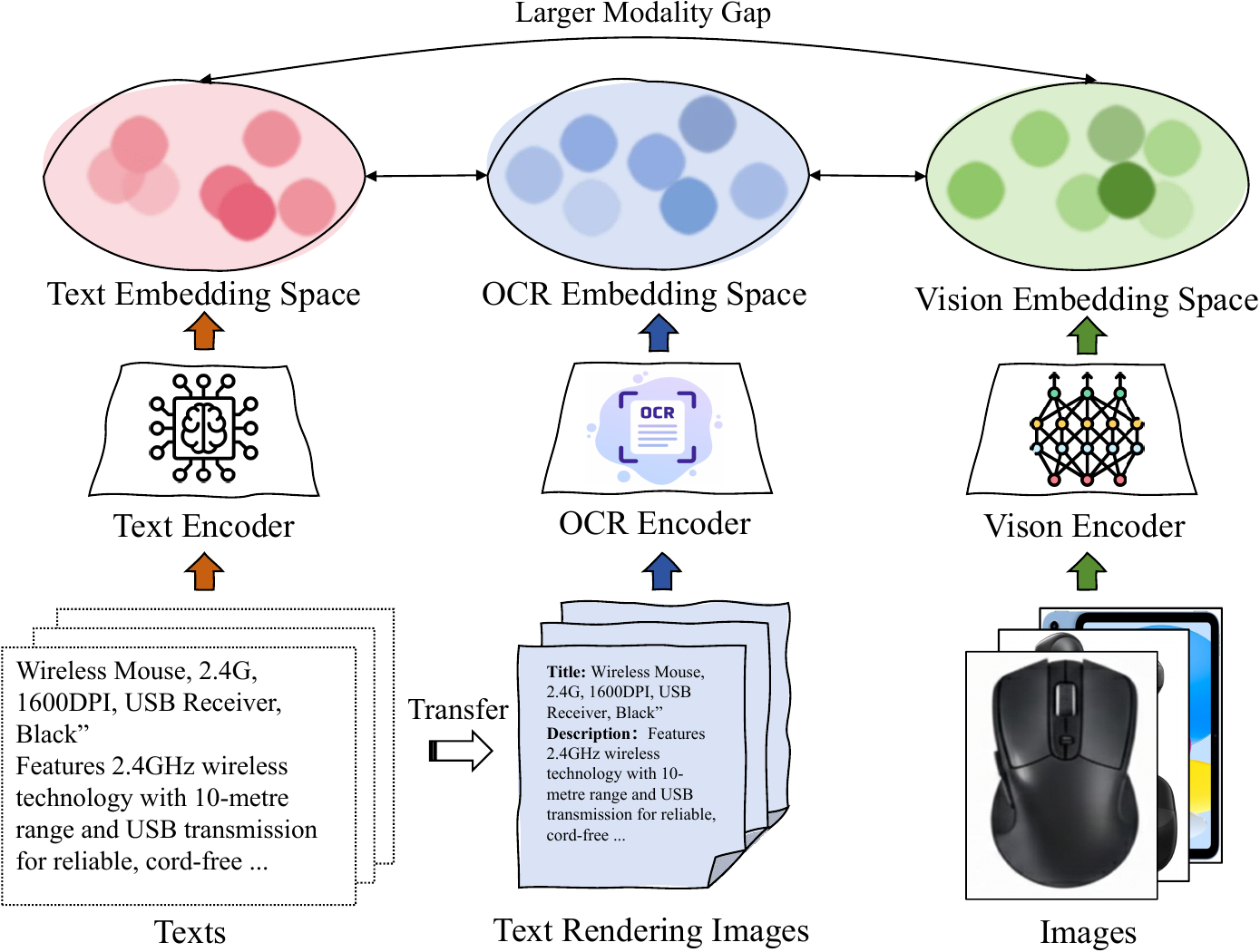}
    \caption{Conceptual illustration of representation spaces induced by different encoders.}
    \label{fig:modality}
    
\end{figure}

\begin{figure*}
    \centering
    \includegraphics[width=0.9\textwidth]{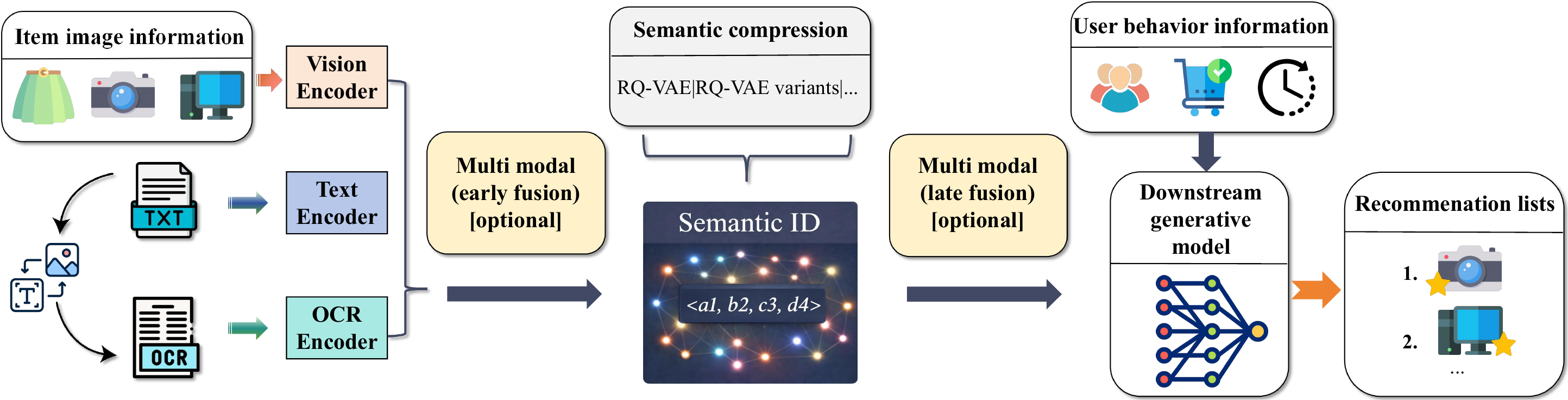}
    \caption{An overview of the Semantic ID-based GR pipeline.}
    \label{fig:pipeline}
    
\end{figure*}

\section{Preliminaries of Semantic ID Learning}
\label{sec:prelim}

\subsection{Overview}
Semantic ID learning enables GR by mapping each item to a sequence of discrete, semantically meaningful codes derived from item content. A typical pipeline consists of three stages: 
(i) extracting modality-specific item representations using pretrained encoders,
(ii) optionally fusing modalities at a chosen stage,
and (iii) learning a semantic tokenizer that discretizes continuous representations into Semantic IDs, which are then consumed by a downstream GR model.
Depending on available side information, the setting can be unimodal (typically text) or multimodal (typically text and images). In the following parts, we mainly focus on introducing the multimodal setting, as the unimodal case can be naturally viewed as a degenerate form obtained by removing the image modality.

\subsection{Generic Item Content Encoding Paradigm}
\label{sec:repr}

Let each item have a textual description $t \in \mathcal{T}$ and (optionally) an image $I \in \mathcal{I}$.
Generally, most Semantic ID methods employ two pre-trained encoders to extract representations from corresponding modality contents.

\noindent\textbf{Text Embeddings.}
A pretrained text encoder
$f_{\text{text}}: \mathcal{T}\rightarrow\mathbb{R}^{d^{\text{text}}}$
maps raw text to an embedding:
\begin{equation}\label{eq_naive_text_emb}
\mathbf{e}_{\text{text}} = f_{\text{text}}(t).
\end{equation}

\noindent\textbf{Image Embeddings.}
A pretrained vision encoder
$f_{\text{img}}:\mathcal{I}\rightarrow\mathbb{R}^{d^\text{img}}$
maps item images to embeddings:
\begin{equation}
\mathbf{e}_{\text{img}} = f_{\text{img}}(I).
\end{equation}

\noindent\textbf{Projection and Normalization.}
To facilitate computation across modalities, Semantic ID methods usually project each modality embedding to a common dimension $d$ with normalization:
\begin{equation}
\tilde{\mathbf{e}}_{m} = \mathrm{norm}\!\left(\mathbf{W}_{m}\mathbf{e}_{m}\right), \quad
m\in\{\text{text}, \text{img}\},
\end{equation}
where $\mathbf{W}_{m}$ is a fixed linear projection (or identity if dimensions already match), and $\mathrm{norm}(\cdot)$ denotes $\ell_2$ normalization.

\subsection{Generic Multimodal Fusion and Token Generation Paradigms}
\label{sec:FusionStrategies}
After obtaining an item’s multimodal representations, the next step is to fuse them. Depending on the stage at which fusion is performed, existing strategies can be broadly categorized into \textbf{early fusion} (feature-level fusion)~\cite{fu2025forge, xu2025mmq} and \textbf{late fusion} (decision-level fusion)~\cite{zhu2025beyond, zhai2025multimodal}. Early fusion integrates representations from all modalities into a single unified embedding, from which Semantic IDs are subsequently generated. In contrast, late fusion first derives Semantic IDs independently for each modality and then combines them at a later stage, after ID generation.

\subsubsection{Early Fusion (Feature-level)}
\label{sec:early_fusion}

Let $\tilde{\mathbf{e}}_i^{(t)}, \tilde{\mathbf{e}}_i^{(\text{img})} \in \mathbb{R}^d$ denote the post-processed textual and image embeddings of item $i$, respectively, where $t$ denotes the textual modality.
The early fusion uses a lightweight gating function:
\begin{equation}
\boldsymbol{\alpha}_i =
g\!\left([\tilde{\mathbf{e}}_i^{(t)} \,\|\, \tilde{\mathbf{e}}_i^{(\text{img})}]\right),
\qquad
\boldsymbol{\alpha}_i \in (0,1)^d ,
\end{equation}
where $g(\cdot)$ is a shallow MLP with a sigmoid activation, and $\|$ denotes vector concatenation.
The fused embedding is computed as
\begin{equation}
\mathbf{z}_i =
\boldsymbol{\alpha}_i \odot \tilde{\mathbf{e}}_i^{(t)}
+
(1-\boldsymbol{\alpha}_i) \odot \tilde{\mathbf{e}}_i^{(\text{img})},
\end{equation}
followed by $\ell_2$ normalization, and $\odot$ denotes element-wise multiplication.
The resulting embedding $\mathbf{z}_i$ is then fed into a Semantic ID generator $\mathrm{SID}(\cdot)$.
\begin{equation}\label{eq_sid_early_fusion}
\mathbf{s}_i = \mathrm{SID}(\mathbf{z}_i),
\qquad
\mathbf{s}_i \in \mathbb{N}^{L}.
\end{equation}
where  $\mathbf{s}_i$ is the fused multimodal Semantic ID tokens, which can be directly used for the downstream tasks. $L$ is the length of a Semantic ID (i.e., the number of tokens in a Semantic ID). 

\subsubsection{Late Fusion (Decision-level)}
\label{sec:late_fusion_concat}
We will introduce three widely used and SOTA late fusion strategies.

\noindent\textbf{Late Fusion A: Independent Semantic ID Concatenation}. As a canonical late-integration baseline, this method preserves complete modality independence during semantic abstraction and defers multimodal interaction until after Semantic ID generation.

For each modality $m \in \{t, \text{img}\}$, a separate Semantic ID generator is trained using the corresponding embedding $\tilde{\mathbf{e}}_i^{(m)}$:
\begin{equation}\label{eq_sid_each_modality}
\mathbf{s}_i^{(m)} = \mathrm{SID}_m(\tilde{\mathbf{e}}_i^{(m)}),
\qquad
\mathbf{s}_i^{(m)} \in \mathbb{N}^{L}.
\end{equation}

Then, these modality-specific Semantic IDs are concatenated directly to form fused Semantic IDs:
\begin{equation}
\mathbf{s}_i =
[\mathbf{s}_i^{(\text{img})} \,\|\, \mathbf{s}_i^{(t)}]
\in \mathbb{N}^{2L}.
\end{equation}

\noindent\textbf{Late Fusion B: Interleaved Variant.}
Based on E.q.~\ref{eq_sid_each_modality}, this fusion strategy aims to fuse modality by cross-arranging different modality Semantic ID tokens:
\begin{equation}
\mathbf{s}_i =
[s_{i,1}^{(\text{img})},\, s_{i,1}^{(t)},\, \ldots,\,
 s_{i,L}^{(\text{img})},\, s_{i,L}^{(t)}].
\end{equation}

\noindent\textbf{Late Fusion C: Modality-aware Tokenization with Alignment Pretraining}.
\label{sec:late_fusion_mgrlf}
The previous late fusion variants treat the multimodal Semantic IDs as a flat sequence, without explicitly marking modality boundaries.
Here, we further consider a modality-aware late-integration setting inspired by MGR-LF++~\cite{zhu2025beyond}.
This setting introduces explicit modality boundaries in the input sequence while keeping embedding extraction and Semantic ID generation unchanged.

For each item $i$, modality-specific Semantic IDs are generated independently and then combined with delimiter tokens:
\begin{equation}\label{eq_align}
\mathbf{s}_i =
[\mathrm{IMG},\ \mathbf{s}_i^{(\text{img})},\ \mathrm{SEP},\ \mathrm{TXT},\ \mathbf{s}_i^{(t)}],
\end{equation}
where $\mathrm{IMG}$ and $\mathrm{TXT}$ denote modality identifiers for images and texts, respectively. $\mathrm{SEP}$ is a separator token that delineates the boundaries between modality tokens.

To introduce controlled cross-modal dependence at the sequence level, we adopt a lightweight alignment pretraining stage by constructing bidirectional translation pairs:
\begin{equation}
[\mathrm{IMG},\ \mathbf{s}_i^{(\text{img})}]
\rightarrow
[\mathrm{TXT},\ \mathbf{s}_i^{(t)}],
[\mathrm{TXT},\ \mathbf{s}_i^{(t)}]
\rightarrow
[\mathrm{IMG},\ \mathbf{s}_i^{(\text{img})}].
\end{equation}
This objective affects only the downstream generator. After alignment pretraining, the model is finetuned on next-item generation based on the multi-modal Semantic IDs $\mathbf{s}_i$ in Eq.~(\ref{eq_align}).

\section{From Text Embeddings to OCR-text Representations}
\label{sec:problem}

Despite rapid progress in GR and Semantic ID learning, most prior work has primarily focused on advancing model architectures and training objectives. In contrast, the problem of how to better represent textual item descriptions in the construction of Semantic IDs has remained largely unexplored. In this work, we explicitly investigate this question and systematically examine its impact on Semantic ID quality and downstream GR performance.

As introduced in Section~\ref{sec_intro} and~\ref{sec:prelim}, existing Semantic ID learning pipelines typically represent item descriptions using embeddings from a pretrained text encoder (e.g., Eq.~(\ref{eq_naive_text_emb})). However, such encoders are optimized for well-formed natural language, while real-world item descriptions are often structured and attribute-centric, containing numerals, units, and abbreviations. Moreover, in multimodal settings, text embeddings are commonly produced by encoder families different from vision encoders, which can lead to semantic and geometric mismatch and complicate cross-modal fusion. 

Motivated by recent progress in vision-centric text modeling, we consider an alternative representation route based on OCR-text. Given an item description $t \in \mathcal{T}$, we first render it into a text image via a rendering function $g:\mathcal{T}\rightarrow\mathcal{X}$:
\begin{equation}
X = g(t),
\end{equation}
where $X\in\mathcal{X}$ denotes the rendered text image (e.g., with a consistent font, size, and layout). We then obtain an OCR-text embedding using a pretrained OCR encoder $f_{\text{ocr}}:\mathcal{X}\rightarrow\mathbb{R}^{d}$:
\begin{equation}
\mathbf{e}_{\text{ocr}} = f_{\text{ocr}}(X).
\end{equation}
By processing text as a visual signal without subword tokenization, OCR encoders can preserve both semantic content and structural cues such as spacing, character shapes, and layout regularities. This representation is more compatible with image embeddings in multimodal fusion. This leads to our central question: Can OCR-text replace the standard text to improve Semantic ID learning for GR, and under what conditions does it help?

To answer this, we conduct comprehensive empirical studies across the settings described in Section~\ref{sec:prelim}, replacing standard text embeddings $\mathbf{e}_{\mathrm{text}}$ with OCR-text embeddings $\mathbf{e}_{\text{ocr}}$ while keeping other components fixed. Based on extensive results, we summarize when and why OCR-text is a better choice for Semantic ID learning.

\section{Experiment}
\subsection{Experiment Setup}

\subsubsection{Datasets and Evaluation Metrics}
\begin{table}[t]
    \centering
    \caption{Statistics of the utilized datasets.}
    \resizebox{0.48\textwidth}{!}{
    \begin{tabular}{c|ccccc} \hline
    \textbf{Datasets}   & \textbf{Users}  & \textbf{Items} & \textbf{Interactions}   & \textbf{Avg.seq.len.} &  \textbf{Avg. word len.}   \\ \hline
    Luxury &1,841  &842  & 18,667  &10.14 & 121.89 \\ 
    Scientific &2,843  &1,549   &19,099& 6.72 &  120.09  \\ 
    Instruments &17,112   &6250   & 136,226  & 7.96 & 103.80\\ 
    Arts & 22,171  & 9,416 & 174,079 &  7.85 & 98.33\\ 
    \hline
    \end{tabular}
    }
    \label{tab:1}
    
\end{table}

We conduct experiments on four commonly used benchmarks~\cite{ni2019justifying} covering various scenarios in GR: \textit{Luxury}, \textit{Scientific}, \textit{Instruments}, and \textit{Arts}. These datasets contain user-item interaction sequences together with item textual and visual descriptions. Table~\ref{tab:1} summarizes the dataset statistics, including the numbers of users, items, interactions, average sequence length, and average text length.

Following standard practice in GR \cite{rajput2023recommender, wang2024learnable}, we evaluate all methods using Recall@K and NDCG@K with $K \in \{5, 10, 20\}$. We follow the standard leave-one-out protocol for sequential recommendation, where the most recent interaction of each user is used for testing, the second most recent for validation, and the remaining interactions for training.
For all the experiments, we run five independent trials with different random seeds and report the average results. 
All improvements are statistically significant with $p<0.05$.

\subsubsection{Implementation Details and Hyperparameter Settings}

All experiments are conducted on a single NVIDIA A40 GPU with 48GB memory using PyTorch~2.5.1 and Python~3.10.
Following common practice in GR~\cite{zheng2024adapting, zhai2025multimodal},
multiple training instances are constructed using a sliding-window strategy with a
maximum history length of 50.

For item representation learning, different pretrained encoders are adopted for different modalities. Specifically, 
textual descriptions are encoded using LLaMA-3.1-8B \cite{touvron2023llama, dubey2024llama}.
Item images are encoded using CLIP with the ViT-L/14 backbone (clip-vit-large-patch14 \cite{radford2021learning}).
OCR-text representations are obtained by rendering item descriptions into images and encoding them with DeepSeek-OCR \cite{wei2025deepseek}.
Unless otherwise specified, all pretrained encoders are kept frozen during Semantic ID learning.

Semantic IDs are generated via residual vector quantization with three quantization layers and a codebook size of 256.
All modality embeddings are projected to 128 dimensions before quantization, and the same Semantic ID generator architecture is used across modalities unless otherwise specified.

For downstream GR, we adopt a T5-style encoder–decoder Transformer architecture, following the TIGER\cite{rajput2023recommender}. For Late Fusion C, we introduce an additional step-based alignment pretraining stage, performed for 2{,}000 steps with a learning rate of $1\times10^{-4}$, prior to task-specific fine-tuning.

During inference, next-item Semantic IDs are generated using beam search with beam size 20 and maximum decoding length 20, and the performance is evaluated using Recall@K and NDCG@K with $K \in \{5,10,20\}$.

\subsubsection{Backbones}
To ensure that our empirical findings are not specific to a particular model design, we conduct experiments on two representative but totally different semantic-ID-based GR backbones:
\begin{itemize}[leftmargin=*]
    \item \textbf{TIGER}~\cite{rajput2023recommender} is a GR framework that predicts the next item by generating discrete Semantic IDentifiers from item side information.

    \item \textbf{LETTER}~\cite{wang2024learnable} learns semantic item tokens via residual vector quantization by adopting an early-fusion paradigm that integrates semantic representations and collaborative filtering signals during tokenization, serving as a multimodal fusion framework for end-to-end optimization in GR.

\end{itemize}

\begin{table*}[]
    \caption{Recommendation Performance of TIGER with Semantic IDs from Text vs. OCR-text.}
    \resizebox{0.8\textwidth}{!}{
    \centering
    \begin{tabular}{c|c|ccccccc}
    \hline
    Datasets & Modality & $\text{Recall}@5$ & $\text{Recall}@10$ & $\text{Recall}@20$ & $\text{NDCG}@5$ & $\text{NDCG}@10$ & $\text{NDCG}@20$ \\
    \hline
    \multirow{3}{*}{Scientific} 
    & Text  &0.0554&0.0795&0.1158&0.0372&0.0454&0.0530\\ 
    & (OCR-text)  &0.0586&0.0840&0.1181&0.0402&0.0481&0.0569\\ 
    & Improve.(\%) & +5.78 & +5.66 & +1.99 & +8.06 & +5.95 & +7.36 \\
    \hline
    \multirow{3}{*}{Luxury} 
    & Text  &0.2628 &0.2826&0.3147 &0.2457&0.2518&0.2608\\ 
    & (OCR-text)  &0.2685&0.2944&0.3178 &0.2482&0.2558&0.2618\\ 
    & Improve.(\%) & +2.17 & +4.18 & +0.99 & +1.02 & +1.59 & +0.38 \\
    \hline
    
    \multirow{3}{*}{Instruments} 
    & Text  &0.0839&0.0946&0.1114&0.0772&0.0807&0.0842\\ 
    & (OCR-text)  &0.0859&0.0970&0.1169&0.0793&0.0829&0.0872\\ 
    & Improve.(\%) & +2.38 & +2.54 & +4.94 & +2.72 & +2.73 & +3.56 \\
    \hline
    \multirow{3}{*}{Arts}
    & Text  &0.0587&0.0678&0.0806&0.0513&0.0541&0.0570\\ 
    & (OCR-text)  &0.0586&0.0680&0.0816&0.0511&0.0543&0.0576\\ 
    & Improve.(\%) & -0.17 & +0.29 & +1.24 & -0.39 & +0.36 & +1.05 \\
    \hline
    \end{tabular}
    }

    \label{tab:unimodal}
    
\end{table*}

\subsection{(RQ1) Text vs. OCR-text Representations in a Unimodal Setting}\label{sec_exp_rq1}

To examine whether OCR-based text representations can serve as a substitute for standard text representations in unimodal Semantic ID learning, we compare recommendation performance using Semantic IDs learned from these two representation forms.

Table~\ref{tab:unimodal} reports the unimodal recommendation performance of TIGER under the two representation choices, 
while LETTER is reserved for evaluating multimodal fusion in subsequent experiments. Notably, OCR-text yields the largest gains on Scientific, whose item descriptions are substantially longer (as shown in Table~\ref{tab:1}) and often follow strong attribute-oriented structures. In this setting, rendering descriptions into images allows OCR-based encoders to better preserve the structural organization of symbols, numerals, and units, which benefits semantic abstraction and discretization. Additionally, Luxury and Instruments exhibit more moderate improvements because their textual metadata is already relatively well-formed and informative, and standard text encoders can capture most semantic information. However, OCR-text representation still outperforms them as it can not only capture the semantics, but also encodes additional structural cues (e.g., formatting and consistent attribute patterns) that are useful for Semantic ID learning. 
The textual description on Arts is shorter and closer to well-formed language, which may be the ideal use case for standard text encoding.
However, according to the results, OCR-text representation can still achieve comparable performance with standard text embedding.

\begin{tcolorbox}[title=Takeaway 1: OCR-text for Unimodal Setting, colback=gray!3, colframe=gray!40]
\small
\textbf{OCR-text is a strong drop-in alternative to the standard text representation for unimodal Semantic ID learning.} OCR-text  clearly outperforms the standard text representation when the item description is lengthy and has attribute-oriented structures. Meanwhile, it can also achieve comparable performance when the description is more like well-formed natural language.
\end{tcolorbox}

\begin{figure*}[t]
    \centering
    \begin{subfigure}{\linewidth}
        \centering
        \includegraphics[width=\linewidth]{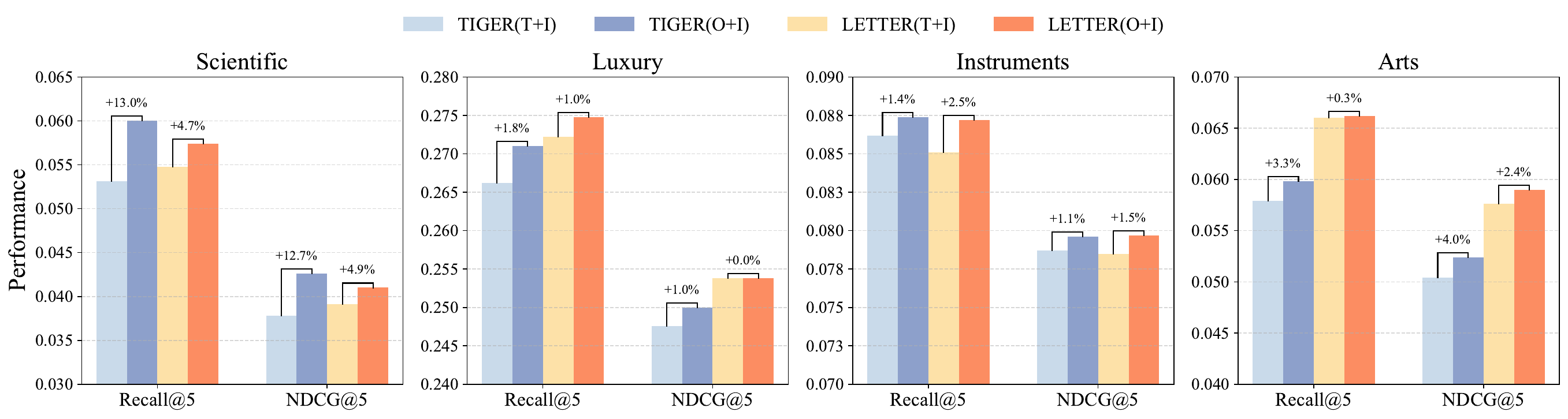}
        \caption{Results under @5 metrics.}
    \end{subfigure}

    \vspace{0.3em}

    \begin{subfigure}{\linewidth}
        \centering
        \includegraphics[width=\linewidth]{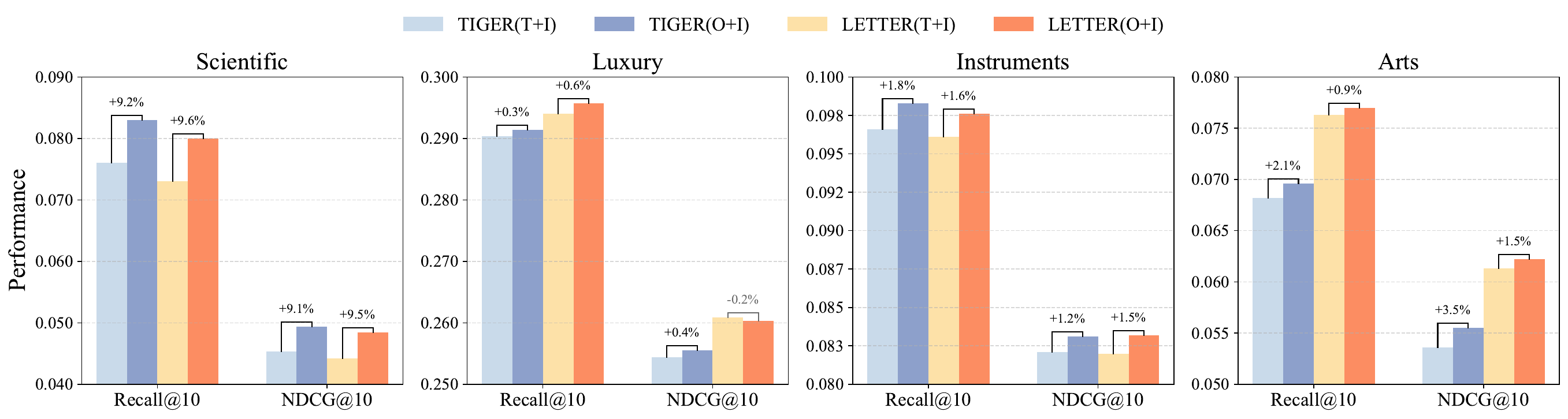}
        \caption{Results under @10 metrics.}
    \end{subfigure}

    \caption{Recommendation Performance of Multimodal Semantic IDs Constructed with the Early-Fusion.}
    \label{fig:earlyfusion}
    \vspace{-10pt}
    
\end{figure*}

\subsection{(RQ2) Text vs. OCR-text Representations in Multimodal Settings}\label{sec_exp_rq2}
As discussed in Section \ref{sec:FusionStrategies}, multimodal Semantic ID learning includes two categories based on how to fuse different modalities: early fusion and late fusion. In this part, we comprehensively investigate each fusion setting to answer whether OCR-text representation can be a better choice.

\subsubsection{Early Fusion Analysis}
We examine whether OCR-text can replace standard text for multimodal Semantic ID learning under an early-fusion paradigm. We denote by \textbf{TIGER(T+I)} and \textbf{LETTER(T+I)} the two backbones using standard text embeddings together with image embeddings, and by \textbf{TIGER(O+I)} and \textbf{LETTER(O+I)} the corresponding variants where the text embeddings are replaced with OCR-text embeddings. Unless otherwise specified, all components other than the textual representation are kept identical, so that any performance difference can be attributed to the representation choice.

Figure~\ref{fig:earlyfusion} summarizes the results on four datasets. On Scientific and Instruments, replacing text embeddings with OCR-text yields significant improvements across all backbones and evaluation metrics. For instance, on Scientific, TIGER(O+I) improves Recall@5 from 0.0531 to 0.0600 and NDCG@5 from 0.0378 to 0.0426 (both more than 12\% gains). On Instruments, OCR-text improves Recall@10 to 0.0983 for TIGER and 0.0976 for LETTER, consistently outperforming their text-based counterparts. 
On Luxury, the improvements from OCR-text are stable but relatively moderate. For example, LETTER(O+I) achieves Recall@10 of 0.2957 and NDCG@10 of 0.2603.
Lastly, on Arts, the improvements from OCR-text representation remain trivial, suggesting limited benefits when item descriptions are short and well-formed.

\begin{tcolorbox}[title=Takeaway 2: OCR-text for Early Fusion Setting, colback=gray!3, colframe=gray!40]
\small
\textbf{OCR-text is an effective drop-in replacement for the standard text representation under early multimodal fusion.}
Moreover, replacing OCR-text in early fusion yields larger relative improvements than in the unimodal setting, because OCR-text representation can produce embeddings that are more compatible with image embeddings and thus easier to fuse compared to standard text representations. Besides, consistent with our unimodal observations, the gains remain strongly tied to description structure: benefits are larger for lengthy, attribute-centric metadata and diminish when descriptions are closer to well-formed natural language.
\end{tcolorbox}

\subsubsection{Late Fusion Analysis}
\begin{figure*}[t]
    \centering
    \begin{subfigure}{\linewidth}
        \centering
        \includegraphics[width=\linewidth]{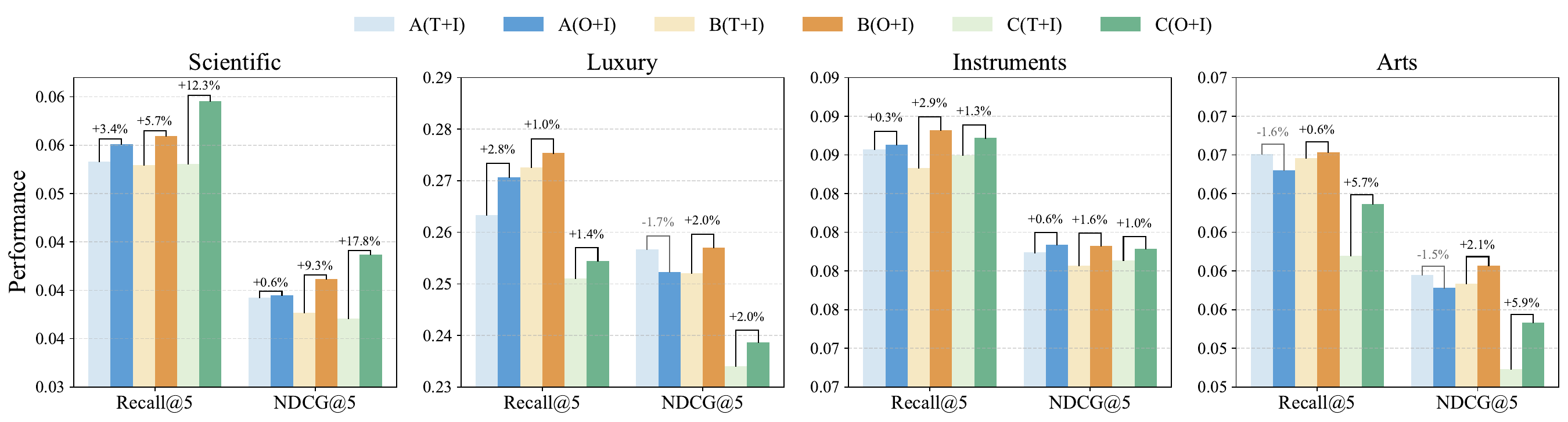}
        \caption{Results under @5 metrics.}
    \end{subfigure}

    \vspace{0.3em}

    \begin{subfigure}{\linewidth}
        \centering
        \includegraphics[width=\linewidth]{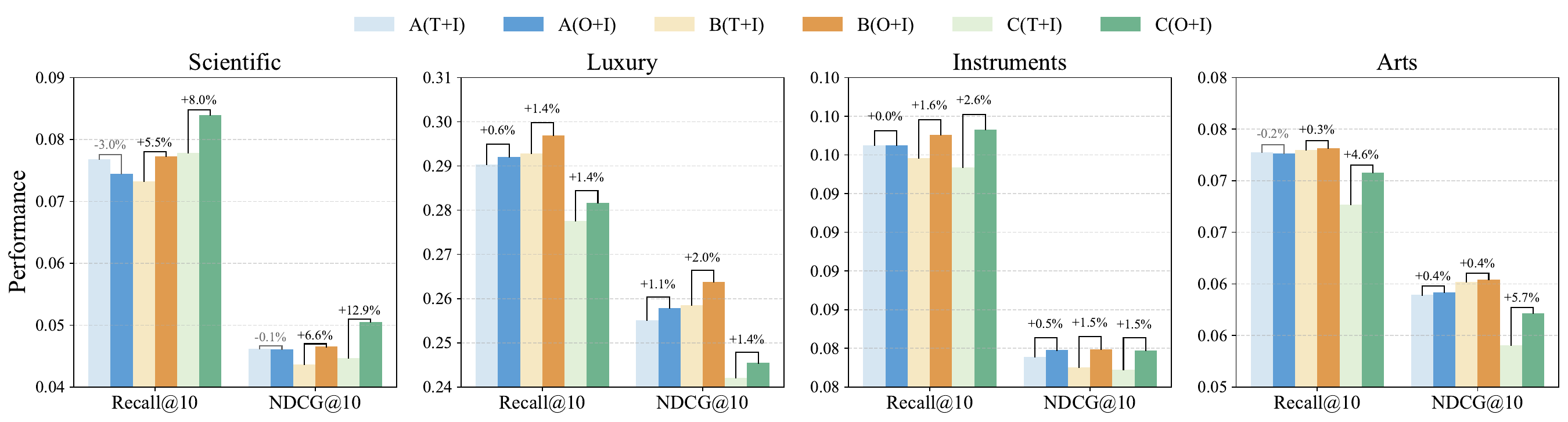}
        \caption{Results under @10 metrics.}
    \end{subfigure}
    \caption{Recommendation Performance of Multimodal Semantic IDs Constructed with Three Late-Fusion Strategies.}
    \label{fig:latefusion}
    \vspace{-10pt}
\end{figure*}

We now study late fusion strategies for multimodal Semantic ID construction. 
We consider three representative strategies introduced in Section~\ref{sec:FusionStrategies}: (A) concatenation, (B) interleaving, and (C) pretrain-based alignment. 
Note that we only adopt TIGER for late fusion investigation, as LETTER is designed for early fusion. 
The results are shown in Figure~\ref{fig:latefusion}.

\noindent\textbf{Late Fusion (A): Concatenation.}
Under concatenation-based fusion, replacing text embeddings with OCR-text achieves overall comparable performance and even yields modest gains on several datasets and metrics. For instance, OCR-text improves Recall@5 on the Scientific, Luxury, and Instruments datasets, with relative gains exceeding 2\% on Scientific and Luxury. The improvement on Instruments is more moderate. On Arts, performance slightly declines, but the magnitude of degradation is negligible. Overall, these results suggest that OCR-text serves as a viable drop-in replacement for standard text embeddings, even under this naive late-fusion strategy.

\begin{tcolorbox}[title=Takeaway 3: OCR-text for Late Fusion (A) Concatenation, colback=gray!3, colframe=gray!40]
\small
\textbf{OCR-text remains a viable drop-in replacement under concatenation-based late fusion,} as it achieves at least comparable performance.
\end{tcolorbox}

\noindent\textbf{Late Fusion (B): Interleaving \& Late Fusion (C): Pretrained Alignment.}
In this part, we compare \textbf{O+I} against \textbf{T+I} under the late fusion strategies (B) and (C). As shown in Figure~\ref{fig:latefusion}, under both Late Fusion (B) and Late Fusion (C), replacing text embeddings with OCR-text yields consistent gains across datasets and metrics, indicating that O+I is generally a better choice for these two strategies. For example, on Scientific, C(O+I) reaches Recall@5 of 0.0596 and NDCG@5 of 0.0437, and the advantage becomes more pronounced at @10 (Recall@10 = 0.0840, NDCG@10 = 0.0505). On Luxury, B(O+I) also shows strong improvements, achieving Recall@5 of 0.2753 and NDCG@5 of 0.2570, with the advantage persisting at @10 (Recall@10 = 0.2968, NDCG@10 = 0.2638). Across datasets, a consistent pattern emerges in the observed gains: the largest improvement occurs on the Scientific dataset, followed by moderate improvements on the Luxury and Instruments datasets, while the Arts dataset exhibits the smallest gain. This trend aligns with the characteristics of item descriptions across the four datasets, which progressively transition from highly attribute-centric metadata to more natural, free-form textual descriptions.

\begin{tcolorbox}[title=Takeaway 4: OCR-text for Late Fusion (B) and (C), colback=gray!3, colframe=gray!40]
\small
\textbf{OCR-text is a strong drop-in replacement for the standard text representation under both interleaving (B) and pretrained alignment (C) late fusion.}
\end{tcolorbox}

\subsection{(RQ3) Robustness Study}
Sections~\ref{sec_exp_rq1} and~\ref{sec_exp_rq2} show that OCR-text can serve as a strong alternative to the standard text representation for Semantic ID learning in both unimodal and multimodal settings. In this section, we further examine the robustness of OCR-text.

\subsubsection{Effect of Rendered Image's Resolution.}
We investigate whether the performance of OCR-text-based Semantic ID learning depends on the quality of the rendered images, since the required image quality directly affects the time and storage costs of rendering and storing text images.

Experiments are conducted on two representative datasets, Scientific and Luxury, which differ markedly in semantic redundancy and information density but have similar text lengths, allowing for a robustness-focused comparison beyond superficial length statistics. For multimodal fusion, we assess robustness using the best-performing fusion strategy for each dataset, that is early fusion for Scientific and late fusion for Luxury.

Table~\ref{tab:compression} reports the relative performance changes when downsampling rendered OCR-text images from $1024\times1024$ to $256\times256$. Across both datasets and fusion settings, the variations are consistently small, mostly within $\pm4\%$. On Scientific, degradations in Recall are negligible, and NDCG remains largely stable, indicating minimal impact on ranking quality.  
On Luxury, reducing image resolution has an even smaller effect: several metrics even show slight improvements in the unimodal setting, and the change in the multimodal setting is at most $1.5\%$.

\begin{tcolorbox}[title=Takeaway 5: OCR-text robustness to Resolution, colback=gray!3, colframe=gray!40]
\small
\textbf{The performance of OCR-text-based Semantic IDs is resilient to aggressive resolution compression} in both unimodal and multimodal settings.
\end{tcolorbox}

\subsubsection{Effect of Different OCR Encoders.}
In previous experiments, we use DeepSeek-OCR to extract OCR-text representations. Here, we examine whether the benefits of OCR-text are tied to a specific OCR backbone by replacing DeepSeek-OCR with two widely used alternatives, Donut-base~\cite{kim2022ocr} and TrOCR-base~\cite{li2023trocr}. Donut is designed for end-to-end document understanding and may better capture higher-level structural regularities, whereas TrOCR emphasizes fine-grained character recognition and glyph-level visual cues. Across all comparisons, we only change the OCR encoder and keep the Semantic ID pipeline, fusion strategy, and downstream GR models unchanged, so that any differences can be attributed to the OCR encoder choice.

Table~\ref{tab:instruments_ocr} shows that all three OCR encoders achieve highly comparable performance, with only minor variations across metrics. For example, Recall@10 ranges narrowly from 0.0961 to 0.0970, and Recall@20 differs by less than 0.004 across encoders, indicating that the overall performance is not sensitive to the specific OCR backbone. Among them, DeepSeek-OCR yields the best results, but the margins are imperceptible in both Recall and NDCG.

\begin{tcolorbox}[title=Takeaway 6: OCR-text robustness to OCR Encoders, colback=gray!3, colframe=gray!40]
\small
\textbf{The advantage of OCR-text is not coupled with a specific OCR encoder.}
Replacing DeepSeek-OCR with TrOCR or Donut leads to highly similar results, demonstrating that OCR-text-based Semantic IDs are robust to the choice of OCR encoders.
\end{tcolorbox}

\begin{table*}[t]
\centering
\caption{Relative performance change (\%) under different OCR rendering resolutions (256 vs. 1024).}
\resizebox{0.8\textwidth}{!}{
\begin{tabular}{c|c|c|cccccc}
\hline
Datasets & Models & Resolution 
& $\text{Recall}@5$ & $\text{Recall}@10$ & $\text{Recall}@20$ 
& $\text{NDCG}@5$ & $\text{NDCG}@10$ & $\text{NDCG}@20$ \\
\hline
\multirow{6}{*}{Scientific}
& Unimodal    & 1024 & 0.0586 & 0.0840 & 0.1181 & 0.0402 & 0.0481 & 0.0569 \\
& Unimodal    & 256  & 0.0570 & 0.0830 & 0.1143 & 0.0405 & 0.0479 & 0.0561 \\
& Rel. Change (\%) & -- 
& -2.73 & -1.19 & -3.22 & +0.75 & -0.42 & -1.41 \\
\cline{2-9}
& Multimodal  & 1024 & 0.0600 & 0.0829 & 0.1164 & 0.0426 & 0.0493 & 0.0571 \\
& Multimodal  & 256  & 0.0584 & 0.0833 & 0.1123 & 0.0409 & 0.0489 & 0.0559 \\
& Rel. Change (\%) & -- 
& -2.67 & +0.48 & -3.52 & -3.99 & -0.81 & -2.10 \\
\hline
\multirow{6}{*}{Luxury}
& Unimodal    & 1024 & 0.2685 & 0.2944 & 0.3178 & 0.2482 & 0.2558 & 0.2618 \\
& Unimodal    & 256  & 0.2715 & 0.2897 & 0.3197 & 0.2489 & 0.2554 & 0.2627 \\
& Rel. Change (\%) & -- 
& +1.12 & -1.60 & +0.60 & +0.28 & -0.16 & +0.34 \\
\cline{2-9}
& Multimodal  & 1024 & 0.2753 & 0.2968 & 0.3317 & 0.2570 & 0.2637 & 0.2717 \\
& Multimodal  & 256  & 0.2741 & 0.2980 & 0.3270 & 0.2550 & 0.2626 & 0.2700 \\
& Rel. Change (\%) & -- 
& -0.44 & +0.40 & -1.42 & -0.78 & -0.42 & -0.63 \\
\hline
\end{tabular}
}
\label{tab:compression}

\end{table*}

\begin{table}[t]
\centering
\caption{Recommendation performance on the \textit{Instruments} dataset using different OCR encoders.}
\label{tab:instruments_ocr}
\resizebox{1.0\linewidth}{!}{
\begin{tabular}{c|ccc|ccc}
\hline
\multirow{2}{*}{Encoder} 
& \multicolumn{3}{c|}{Recall} 
& \multicolumn{3}{c}{NDCG} \\
& @5 & @10 & @20 & @5 & @10 & @20 \\
\hline
Donut-base 
& 0.0846 & 0.0961 & 0.1131 
& 0.0784 & 0.0820 & 0.0861 \\

Trocr-base-printed
& 0.0850 & 0.0964 & 0.1157 
& 0.0773 & 0.0810 & 0.0849 \\

DeepSeek-OCR 
& \textbf{0.0859} & \textbf{0.0970} & \textbf{0.1170} 
& \textbf{0.0793} & \textbf{0.0829} & \textbf{0.0872} \\
\hline
\end{tabular}
}

\end{table}

\subsection{Case Study:  Dataset-Specific Effects of OCR-text Representations}
\label{sec:casestudy}
\begin{figure}
    \centering
    \includegraphics[width=1.0\linewidth]{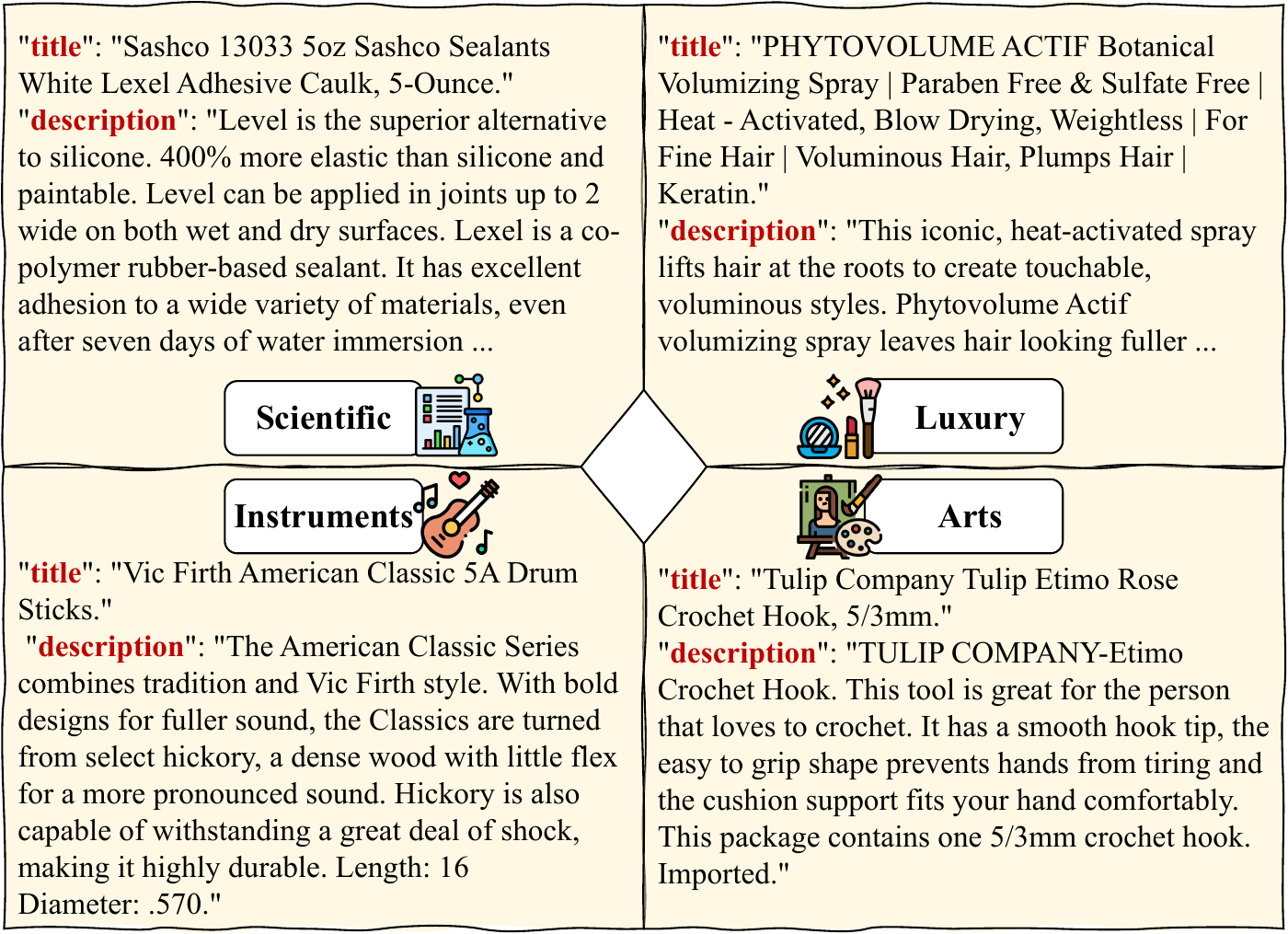}
    \caption{One representative item description from each dataset: Scientific, Instruments, Luxury, and Arts.}
    \label{fig:case}
    
\end{figure}

To better contextualize the dataset-dependent trends observed in Sections~\ref{sec_exp_rq1} and \ref{sec_exp_rq2}, 
Figure~\ref{fig:case} shows a case study by inspecting representative item descriptions from the four datasets.

As we can see, textual item descriptions from Scientific and Instruments are typically specification-oriented and attribute-dense, containing abundant numerals, units, and compact attribute lists (e.g., dimensions, tolerances, materials, and technical parameters). Such metadata relies heavily on symbolic cues rather than fluent natural language. When rendered into images, OCR-text can preserve fine-grained character information while making structural regularities explicit, such as consistent separators, spacing patterns, and clear attribute boundaries. These observations align with the consistent gains of OCR-text on both Recall and NDCG in these datasets.

In contrast, descriptions of Luxury and Arts are more narrative- or marketing-oriented, often dominated by subjective adjectives and stylistic expressions with fewer explicit attributes or rigid structures. In this case, rendering text into a visual form introduces limited additional information beyond what standard text encoders already capture, which is consistent with the smaller improvements on Luxury and the near-neutral differences observed on Arts.

\begin{tcolorbox}[title=Takeaway 7: Why OCR-text Representation Helps, colback=gray!3, colframe=gray!40]
\small
Compared to the standard text representation, \textbf{OCR-text representation gains mainly come from capturing structural cues in item descriptions.}
When descriptions are more attribute-based (dense numerals, units, and key-value style patterns), OCR-text representation not only encodes semantic information but also captures clearer structure and yields larger improvements.
When descriptions are closer to well-formed natural language, OCR-text encodes similar semantics as standard text embeddings, so the gains become smaller.
\end{tcolorbox}

\section{Related Work}
\subsection{Semantic ID in Generative Recommendation}
Recent advances in GR introduce Semantic IDs as a compact and expressive interface between item content and autoregressive sequence models.
TIGER~\cite{rajput2023recommender} is a pioneering work that reformulates sequential recommendation as a generative retrieval problem by learning discrete Semantic IDentifiers via vector quantization, enabling efficient decoding independent of the item corpus size.
VQ-Rec~\cite{hou2023learning} further demonstrates that vector-quantized item representations improve transferability and robustness in sequential recommendation.

Subsequent research has focused on enhancing the structure, objectives, and stability of Semantic ID learning. LETTER~\cite{wang2024learnable} and SETRec~\cite{lin2025order} respectively improve tokenization quality through residual quantization with collaborative alignment and order-agnostic identifier design, while HiD-VAE~\cite{fang2025hid} introduces hierarchical and disentangled Semantic IDs to improve interpretability and reduce collisions. In parallel, several works~\cite{liu2025generative, ye2025dual, li2025semantic} explore alignment-centric and end-to-end frameworks, including generative tokenization with end-to-end optimization, dual-aligned Semantic IDs for industrial systems, and multi-stage semantic alignment between recommendation models and language models.

Beyond identifier learning, recent studies further investigate scalability and inference efficiency in Semantic ID-based GR.
Building on this line of work, RPG~\cite{hou2025generating} improves decoding efficiency through parallel Semantic ID generation, while hybrid approaches such as COBRA~\cite{yang2025sparse} combine sparse Semantic IDs with dense representations to mitigate information loss.

Existing Semantic ID-based GR methods implicitly assume standard text embeddings as a fixed input, and focus on optimizing identifier structures or tokenization.
We are the first to systematically study how replacing text embeddings with OCR-based visual text representations affects Semantic ID learning.

\subsection{Multimodal Recommendation}

Multimodal side information, such as textual descriptions and product images, has long been leveraged to enrich item representations and improve recommendation quality.
Early representative methods include VBPR~\cite{he2016vbpr}, which incorporates visual features into matrix factorization, and a series of graph-based models (e.g., LATTICE~\cite{zeng2021pan}, DualGNN~\cite{wang2021dualgnn}, and MVGAE~\cite{yi2021multi}) that capture modality-aware item relations and user preferences through structured message passing.

More recently, GR has shifted toward learning discrete Semantic IDentifiers derived from item content. MMGRec~\cite{liu2024mmgrec} represents one of the earliest attempts to integrate multimodal content and collaborative signals into Semantic ID generation by quantizing fused representations with a graph-based RQ-VAE.
FORGE~\cite{fu2025forge} provides one of the first large-scale industrial benchmarks for forming Semantic IDentifiers from multimodal features, and systematically studies optimization strategies and evaluation metrics for Semantic ID generation.
MMQ~\cite{xu2025mmq} further advances multimodal Semantic ID learning by introducing a shared--specific mixture-of-quantization tokenizer that explicitly balances cross-modal synergy and modality-specific uniqueness, and adapts Semantic IDentifiers to downstream behavioral signals.

In parallel, several works investigate multimodal GR from a model-centric perspective.
Zhu et al.~\cite{zhu2025beyond} presents the first systematic study of multimodal GR, revealing the sensitivity of Semantic ID generation to modality choices and fusion strategies, and highlighting the limitations of naive early- and late-fusion designs.
Building upon this observation, MQL4GRec~\cite{zhai2025multimodal} proposes to translate multimodal item content into a unified quantitative language, enabling cross-domain and cross-modality knowledge transfer through a shared discrete vocabulary.

Prior multimodal GR methods remain embedding-centric, fusing multimodal signals in continuous space and using Semantic IDs as fixed downstream interfaces.
In contrast, we explicitly revisit the text representation used for Semantic ID learning by replacing standard text embeddings with OCR-based visual text representations.

\subsection{Visual Text Representations}
Recent work has explored visual text representations by converting raw text into images and processing them with vision or OCR-capable models. Early studies such as PIXEL and PIXAR\cite{tai2024pixar} demonstrate that pixel-based language modeling can bypass subword tokenization, improving robustness to character-level noise and multilingual variation, while even enabling autoregressive generation directly in pixel space. More recent approaches leverage visual text primarily as a token-efficient interface for large models. Glyph\cite{cheng2025glyph} renders long documents into compact images to scale context length, achieving substantial token compression without degrading performance, while Text-or-Pixels\cite{li2025text} shows that feeding long contexts as images can nearly halve decoder token usage in multimodal LLMs. From a tokenization perspective, SEETOK\cite{xing2025see} and VIST\cite{xing2025vision} treat rendered text as a vision-centric alternative to subword tokenization, aggregating character shapes and layout cues into compact visual tokens and improving efficiency and robustness. 

While visual text representations have been explored in language modeling and multimodal understanding, their role in Semantic ID learning for GR has not been examined.
This work presents the first empirical study integrating OCR-based visual text into Semantic ID learning.

\section{Conclusion}
This paper presented a systematic empirical investigation into the impact of text representations on Semantic ID learning for generative recommendation. Moving beyond the prevailing text-centric assumption, we examined whether OCR-based visual text representations can serve as a drop-in replacement for the standard text representation when constructing Semantic IDs for generative recommendation, in both unimodal and multimodal settings. Across a diverse set of datasets, model backbones, and fusion strategies, our results demonstrate that OCR-text consistently matches or surpasses standard text representations for Semantic ID learning. The advantages are most pronounced on datasets characterized by dense, attribute-centric descriptions, where OCR-text more effectively preserves numerals, units, and structural regularities that are critical for discretization. In multimodal recommendation, OCR-text also exhibits stronger compatibility with item image embeddings, frequently yielding improvements under both early and late fusion strategies. Moreover, robustness analyses show that OCR-text-based Semantic IDs remain resilient to practical perturbations, including aggressive rendering resolution compression and variations in OCR encoders, underscoring their suitability for realistic deployment scenarios.

Taken together, these findings highlight the effectiveness of vision-centric text representations in Semantic ID learning. We hope this work motivates future research to re-examine representation assumptions in generative recommender systems and to explore broader vision-centric tokenization and fusion paradigms for multimodal semantic abstraction.


\bibliographystyle{ACM-Reference-Format}
\bibliography{mybib}

\appendix

\end{document}